\documentstyle{aa}
\begin{document}

\thesaurus{11(11.07.1)}

\title{The use of luminosity effects to calculate scaling
regimes in the galaxy distribution}

\author{  Reuben Thieberger \inst{1}}

\institute{ Physics Department, Ben Gurion University, Beer Sheva, 
84105 Israel}

\mail{ thieb@bgumail.bgu.ac.il}

\date{Received 30 April 2002 / Accepted 7 June 2002 }

\maketitle

\mark{R. Thieberger: The use of Luminosity Effects for calculating
Scaling regimes in the galaxy distribution}

\begin{abstract}
This work is a continuation of a previous publication { Thieberger 
\&  Spiegel} concerning the use of two dimensional catalogues
in unraveling the large scale distribution of galaxies. Since the
data in two-dimensional catalogues are more plentiful than those in 
three dimensions, the luminosity method can be helpful in using
two dimensional data to unravel the nature of the three dimensional
distribution. In this work we analyse the Cfa catalogue and 
demonstrate that the results indicate a distribution of two dimensions 
in the nearer range and three dimensions in the farther range. 
The possibility of a different regime for very short distances is 
discussed.  \\

\end{abstract}

\section{Introduction}

Standard cosmology is based on the assumption that the universe is 
spatially homogeneous, at least on scales sufficiently large to 
justify its approximation by a
 Friedmann-Lema\^itre-Robertson-Walker (FLRW) model. 
The high \linebreak isotropy measured in the cosmic microwave 
background radiation (CMBR) is usually taken as strong evidence 
in support of this hypothesis. \\

The structures seen in galaxy catalogues - groups, clusters and 
superclusters, distributed along voids, filaments and walls -
 are not viewed as contradicting this Principle, as the common
opinion is that the scales on which the universe is assumed 
to be homogeneous are much larger than those subtended by 
these structures.  However, the consensus on a 
homogeneous feature of structures, even on very large scales, 
has never been complete (see e.g. Pietronero ,1987).  \\

In more recent times, three-dimensional galaxy catalogues have been 
made available, they support the findings of approximate 
self-similarity in the large-scale distribution of galaxies
within prescribed scale intervals (Provenzale, 1991). Since the
data in two-dimensional catalogues are more plentiful than those in three
dimensions and are likely to remain so for some time (Cress, et al., 1996),
it seems worthwhile to discuss further the ways in which the
two-dimensional data may be used to unravel the nature of the
three-dimensional distribution. The basic issue in analyzing a 
two-dimensional catalogue goes back to work on stellar statistics 
(Trumpler and Weaver, 1953).The significant point in the stellar 
studies was the effect of the luminosity distribution of the 
observed objects on the projected distributions. \\ 

The general outline of this issue, which is the main concern of our
present attempt has been given previously (Thieberger and Spiegel). 
Just as in the abovementioned publication, we shall leave out of 
the account some other features of the galaxy case that remain to be 
considered in detail, namely source evolution and cosmological 
effects (Spedalere and Schucking, 1980; Ribeiro, 1995 and 
Celerier and Thieberger, 2000). \\

In this study we limit ourselves to a three dimensional catalogue so as
to be able to evaluate the results. Therefore we chose the cfa catalogue, 
which has been widely studied, and therefore where we can perform all necessary 
comparisons so as to understand more precisely our approach. \\

One of our motivations for the present study comes from our
interest in trying to distinguish the different scaling regimes
that may exist in the galaxy distribution, though on different
length scales.  It has been suggested (Murante et al., 1998) 
that three scaling regimes may be discerned in the
distribution of galaxies: on the smallest scales the results are
consistent with a distribution of density singularities; in
the intermediate range, there is scaling behaviour suggestive of
flat structures such as Zeldovitch (1970) favored; on the largest
scales, the data indicate a homogeneous distribution of galaxies
with nonfractal behavior.  This categorization is based on the
results of an analysis of three-dimensional catalogues. However,
it has to be admitted that those data may as yet not be adequate
to clearly decide such issues and that is why we wish to
consider whether we may reliably use two-dimensional catalogues for
immediate needs. Because of statistical problems, we find it at the
moment difficult to study the close region behaviour, therefore we will
limit ourselves in this study to the two other regions namely,
the two dimensional region and the three dimensional one. 

Before describing our calculations, we devote the two next sections 
to a short description of the correlation integral method and 
to a brief outline of the influence of luminosity effects by 
imposing a Schechter luminosity function on the objects of 
study and attempt to see how this influences the distribution of 
the projected objects, or rather of their apparent magnitudes.
Then we devote a section to a model catalogue so as to help
us understand the results of the observational data.

\section{The correlation Integral method.}

Calculating dimensions for galaxies enters into difficulty 
for the cases when the distribution is fractal (Coleman and 
Pietronero 1992). A robust characterization of the structures 
of point sets is given by the {\it correlation integral} (Grassberger 
and Procaccia, 1983), defined as:  
 
\begin{equation}
C_2(r)= {1\over {N^{\prime}(N-1)}}\sum_i \sum_{j\ne i}
\Theta(r - |{\bf X}_i - {\bf X}_j|)
\label{cor}
\end{equation}
where $\Theta$ is the Heaviside function.The inner summation is over 
the whole set of $N-1$ galaxies with coordinates ${\bf X}_j$, 
$j\ne i$, and the outer summation is over a subset of $N^{\prime}$ 
galaxies, taken as centers, with coordinates ${\bf X}_i$. By 
taking only the inner $N^{\prime}$ galaxies as centers we allow 
for the effect of the finiteness of the sample (see e.g. 
Sylos Labini et al., 1998). 

We may interpret $C_2(r)$ as ${\cal N}(r)/N$ where ${\cal N}(r)$ is the
average number of galaxies within a distance $r$ of a typical galaxy in
the set.  As $r$ goes to zero, $C_2$ should go to zero and, for general
distributions, we express this conclusion as $C_2 \propto r^{D_2}$.
The exponent $D_2$ is called the correlation dimension
and it is necessarily $\le 3$ for an embedding space of dimension three.
When $D_2$ is not an integer, the distribution is called fractal
(Mandelbrot, 1982). 

In Eq.(\ref{cor}) we have a double summation. In cer\-tain cases, for 
example when we have a pencil beam, this inner region has to be 
very small and so sometimes one has to use just the
distances from a single point. This of course results in a strong
deterioration of the statistics. For comparison, we will perform
such calculations  and we will call it the "pencil beam method" 
(the dimension obtained will be denoted by $D_{PB}$,comparing it
to the usual "Grassberger Procaccia method" (with dimension $D_{GP}$).

\section{Luminosity effects.}

In considering the distribution on the celestial sphere of galaxies 
or of radio sources the role of the intrinsic luminosities of 
the observed objects must be allowed for. Such effects appear 
in $N(>f)$, the number of galaxies detected whose flux density 
exceeds a given value, $f$ (Peebles, 1993). As is done in 
Thieberger and Spiegel, we assume that the luminosity function 
of the objects in our study is independent of position and of time.

A galaxy at distance $r$ is bright enough to be seen at flux 
density,$f$, if its luminosity exceeds $L$. Using a characteristic 
luminosity $L_*$, for normalization one can express the mean 
number of galaxies per steradian brighter than $f$ by (Peebles, 1993):
\begin{equation}
N(>f)=const. \int^{\infty }_0 r^{D-1} dr \psi (4 \pi r^2f/L_*)
\label{d2}
\end{equation}
$D$ being the dimension of the galaxy distribution, whether fractal
or regular. $\psi (x)$, is proportional to the probability that a
galaxy at distance $r$ is included in the catalogue. For a Schechter 
model (Schechter, 1976),
\begin{equation}
\psi (x)=\int^{\infty }_x w^{\alpha } \exp^{-w} dw
\label{d3}
\end{equation}
Here $\alpha =-1.07$ and $x=L/L_*$.

Lets consider our specific problem. We assume that (see 
Murante et al., 1998) at short distances we have a dimension 
D=2 ("Zeldovitch panckakes") and at large distances we have 
homogenous distribution (i.e.  D=3). In this picture we have
ignored the very close region where we probably have D=1 (the 
singularity picture).  Then we can modify Eq.(\ref{d2}), 
and write:
\begin{equation}
N(>f)\sim f^{-1} \int^{r^*_0}_0 y dy \psi (y^2) +
f^{-1.5} \int^{\infty}_{r^*_0} y^2 dy \psi (y^2) 
\label{d4}
\end{equation}
Here $y=4 \pi r^2f/L_* $, and $r^*_0 = r_0 \sqrt{4 \pi f/L_*}$.
$r_0$ is the value at which we change the dimension from two
to three. It is easy to see that for the limit of small f the 
value of the slope $\log N(>f)$ vs. $\log f$ tends to $D=3$, and 
for large $f$ it tends to $D=2$. 
We used this equation to evaluate numerically the dimension, 
for the whole range of $f$, by calculating an array of values 
of $f$ vs. $N(>f)$, making a numerical integration of Eq.(\ref{d4}). 
 We obtained then the dimension by a least squares fit. 
This analysis resulted in $D=2.9$ for very small $f$,
and going fast towards the value of $D=2.$ , for large $f$. 
These results are of course quite obvious,the only point we 
wanted to clarify is the extension of the transition 
zone between 2. and 3. dimensions. We found that this transition 
zone is larger than the zone where f tends to zero and that quite
small values of $f$  already give a dimension $D=3.$, i.e.
are already behaving as very large $f$.
In the next section we will perform a number of model catalogue
calculations in order to study the relations between the results
obtained by the luminosity method versus the integral correlation
method. We will also compare that to the 'beam pencil method' as
was explained in the previous section. The reason is that we wish
to stress that in the "luminosity method" we are forced to consider
only the luminosities relative to the observer, i.e. just for one
galaxy.

\section{The model catalogues }
In order to understand the results which we obtain from an
observational catalogue, we constructed a number of representative
model catalogues.  We took $N_1$ points in a box 0 to 500 
(arbitrary units), and around each point on a plane we put 
$N_2$ points at distances up to $PL$ units, in each direction 
on the plane. To each point we attributed a 'luminosity' 
obtained according to the Schechter function distribution 
(Eq. (\ref{d2})).  In most of the catalogues we  considered 
the cases were we had in our box around 20000 points 
($N_1 \times N_2$ minus the points which fall outside the 
box limits). This value was chosen so as to be of the same 
order of magnitude as the Cfa catalogue. 
In this way we can examine the reliability of the results
for the real catalogue. Just for the sake of learning about the
improvement obtained for larger catalogues, we checked also a few
cases of a larger amount of points. In the following table
we compare the results to the ones obtained by the use of 
correlation integrals (Grassberger and Procaccia, 1983).
In the table we compare our results also to the pencil beam method.
The dimensions are denoted by $D_{GP}$ and $D_{PB}$ appropriately.
In this table we give also the results obtained by performing
the dimension analysis  via the luminosity effect.  
Using the $log(f)$ versus $log(N(>f))$ results, we calculate
the appropriate dimensions. The result is denoted by $D_L$ in
the table.

In addition to the main purpose of the table, which is the
comparison beween the different methods, for a variaty of
cases, we also obtain an indication of the errors for the
different models. The cases of pure two or three dimensional
models (the last four lines in the table) give good agreement
and when the number of points is increased (see last line)
the errors even for the luminosity case start to decrease
considerably. The main reason for the larger overall errors
in the top five lines is the consequence of our model where
some points 'see' more three dimensional points then others
depending on their location on the planes. This problem, being
an artifact, should be less pronounced in the true 
catalogues.

\begin{table}
\caption{Dimension calculation for different model catalogues.}
\begin{tabular}{cccccc}
\hline
$N_1$&$N_2$&$PL$&$D_{GP}$&$D_L$&$D_{PB}$ \\
\hline
150&150&50&2.3 $\pm $ 0.1&2.5 $\pm $ 0.1&2.5 $\pm $0.2\\
300&150&50&2.6 $\pm $ 0.1&2.6 $\pm $ 0.1&2.8 $\pm $0.2\\
300&300&50&2.6 $\pm $ 0.0&2.6 $\pm $ 0.1&2.6 $\pm $0.2\\
6&4800&200&2.2 $\pm $ 0.2&2.3 $\pm $ 0.2&2.4 $\pm $0.3\\
3700&6&50&2.95 $\pm $ .05&2.9 $\pm $ 0.1&3.0 $\pm $0.1\\
1&20k&250&2.00 $\pm $ .00&2.00 $\pm $ .04&2.01 $\pm $.03\\
20k&1&.1&3.00 $\pm $ .00&2.9 $\pm $ 0.1&2.98 $\pm $.05\\
80k&1&.1&3.00 $\pm $ .00&2.9 $\pm $ 0.1&3.00 $\pm $.03\\
320k&1&.1&3.00 $\pm $ .00&2.94 $\pm $ .03&2.99 $\pm $.02\\
\hline
\end{tabular}

\end{table}

We  wish to remark here that
there are problems concerning the validity of the results
both for very small $f$ and for large $f$. For small $f$
the finite size of our box results in the limit being $f$
dependent, which is an artifact. For large $f$ only a
smaller and smaller fraction of the points reach the
"observer", resulting in bad statistics.
So we check the average dimension  over the range of $f$
where we thought the results were reliable.  \\

\section{Observational Catalogue}

In this section we describe the dimensions that we obtain
based on the three dimensional data of the Cfa Catalogue of
redshifts (Huchra et al. ,1995), which is a compilation
from several sources. A previous analysis using the standard
method, described in section 2, was performed, for an older
catalogue, by Provenzale et al.,(1997).\\

We wish to compare the results 
of the correletion in\-te\-gral method and the lumi\-nos\-ity method.
The correlation integral method gives for the near galaxies
$D=1.9$ and for the far galaxies $D=2.6$. The luminosity
method gives for the  mostly near (as explained in the previous 
section) $D=2.1$, and for the mostly far ones: $D=2.8$.

In these cases we do not have a way to put error bars on the
values. So to get a feel for the possible error, we attribute 
to each galaxy a luminosity obtained in the same manner as in 
the model catalogue, instead of using the values of the 
luminosities in the real Cfa catalogue. We chose nine different 
sets of random numbers. In this way we  could obtain
the error on the average and obtained an error of about
$ \pm 0.2$. These are encouraging results demonstrating that 
although the errors for the luminosity method are somewhat larger 
than the ones for the correlation integral method, they are in
a reasonable range . Of course if one has  a large
two dimensional catalogue and a relatively much smaller three
dimensional catalogue (as is at the moment the case) then the
larger number of galaxies in the two dimensional catalogue will 
make the errors smaller then the one obtained from analysing a 
red shift catalogue.

\section{Discussion and conclusion}

Our calculations show that we obtain the same kind of results
both for the correlation integral method and the luminosity
method. We tried to estimate the possible errors on the results
and it seems to us that they are somewhat larger for the luminosity
method. So if we have a three dimensional catalogue most probably
the correlation integral give a more reliable result, but as the
two dimensional catalogues are much larger, most probably the
luminosity method is more reliable.

{\it Aknowledgements.}  
 The author wants to thank Professor EA Spiegel for many valuable
discussions concerning this work.

\end{document}